# Simulation Modelling of Inequality in Cancer Service Access

Ka C. CHAN [a,1], Ruth F. G. WILLIAMS [b,c], Christopher T. LENARD [b], and Terence M. MILLS [b]

[a] *School of Business, Education, Law and Arts, University of Southern Queensland*
*Springfield Central, QLD 4300, Australia*
[b] *Department of Mathematics and Statistics, La Trobe University*
*Bendigo, VIC 3552, Australia*
[c] *Graduate School of Education, The University of Melbourne*
*VIC 3052, Australia*

**Abstract.** This paper applies economic concepts from measuring income inequality to an exercise in assessing spatial inequality in cancer service access in regional areas. We propose a mathematical model for accessing chemotherapy among local government areas (LGAs). Our model incorporates a distance factor. With a simulation we report results for a single inequality measure: the Lorenz curve is depicted for our illustrative data. We develop this approach in order to move incrementally towards its application to actual data and real-world health service regions. We seek to develop the exercises that can lead policy makers to relevant policy information on the most useful data collections to be collected and modeling for cancer service access in regional areas.

## Introduction

Access to treatment is an interesting topic in cancer services research [1]. Cancer treatment often involves three major types of therapy: surgery, chemotherapy and radiotherapy. Each type of therapy can present a different accessibility issue in regional areas as different travel patterns may be needed in order to utilise cancer therapy. "Accessibility" is on the minds of cancer service users and providers as service users can experience spatial or geographic inequalities [2]. There is an established literature in economics on measuring inequality, although the variable that is mostly the subject of inequality measurement is income [3–4]. People with cancer can experience economic inequalities in cancer treatment beyond income inequality. For cancer treatment inequality, some of the relevant questions for measuring inequality can change from those asked about income inequality [5]. This shift in research focus is in part because many countries have health insurance schemes now and they ease burdens borne by cancer service users and providers from the price/s paid and received for cancer services. Another type of inequality burden that cancer service users can experience is physically accessing services for cancer treatment [2].

---

[1] Corresponding Author: Ka C. Chan; E-mail: kc.chan@usq.edu.au

This paper applies some ideas from measuring inequalities in economics to assessing spatial inequalities in cancer service utilisation. Some of the inequality measures often applied, such as depicting a Lorenz curve or calculating a Gini coefficient or Atkinson Measure [6] are relevant measures initially, but we wish to focus on the variable that we seek to measure, viz. utilisation/cancer incidence. In cancer service accessibility, there is a need for a particular commodity that is a service, and services may not be delivered to shops and supermarkets where they can be bought, as is the case when people purchase goods like bread, apples, shoes, cars etc., all of which are final products. Services can often be different from goods: people in need of a service need to travel to the point of service production. With cancer treatment, the geographic distribution of cancer service production locations may not perfectly correspond to the geographic pattern of cancer incidence. Thus, spatial inequalities in service access may develop.

In the case of cancer in regional areas, the level of service utilisation in an area relative to the cancer incidence for that area may be a phenomenon warranting better measurement. Data on incidence exist, and there are also publicly available data on service utilisation in regional areas. These are hospital separation data. With existing data, we examine how these data can inform policy, and ask what other data may need to be collected and which variables need to be measured and modelled to inform policy makers usefully. The ultimate purpose is that policy makers are better informed about the problem of spatial inequality in cancer service access in regional areas.

The aim of our paper is to develop an exploratory model of spatial inequality in access to cancer services using the measure/s that we investigate, which are developed from available data. The measure of cancer service access used here is the ratio of separation/incidence and we seek to explore this measure for a single health service region in Victoria, Australia. In this exploratory study, we measure spatial inequality in accessing chemotherapy only, as the initial exercise with the purpose of guiding further studies. Future work can model the data publicly available for all three main cancer therapies by health service regions, viz. chemotherapy, radiotherapy, and surgery.

The Victorian Department of Health and Human Services delineates five rural health service regions for the State of Victoria: Hume, Gippsland, Barwon South Western, Loddon Mallee and Grampians. These five regions exist in addition to the metropolitan health service areas. Each of these regions is composed of several Local Government Areas (LGAs). We develop our modelling and simulation tools on illustrative data for a hypothetical health service region having 13 LGAs. The initial exercise here will not incorporate data on the important distinction that exists within the institutional framework of the Australian health system. viz. between publicly- and privately-funded hospital services and where these services are located. This distinction affects the prices paid and received in regard to health service utilisation in Australia and, in turn, the spatial distribution of cancer service accessibility; however, for the hypothetical exercise here, modelling data on that distinction is not vital.

**1. Modelling Spatial Inequality in Access to Cancer Services**

*1.1. Chemotherapy Service Utilisation Data*

The "separation" is a measure of service utilisation. A separation from hospital for chemotherapy occurs when a patient leaves a hospital. One patient having a course of chemotherapy may be involved in several separations. The following data are collected and publicly available for chemotherapy: (1) Incidence (number of new cases) of all cancers in each LGA in a year. (2) Number of same day separations for chemotherapy for the year generated by patients in the LGA. (3) Number of patients from each LGA who used chemotherapy during the year.

*1.2. Modelling Spatial Inequality Among Local Government Areas (LGAs)*

The core of this paper borrows ideas from research in inequality. Cowell states that three features of the model must be defined [3]:

- The unit of analysis such as a single person, the nuclear or the extended family.
- The attribute to be measured such as income, wealth, land-ownership, voting strength or, as in this case, access to cancer treatment.
- A method of representation or aggregation of the allocation of the attribute among the persons in a given population.

The unit of analysis in this study is a person diagnosed with cancer needing chemotherapy, but the available data are grouped at LGA level. The attribute subject to the inequality being measured is spatial access to chemotherapy. The method for representing this inequality is to be a Lorenz curve only. Reporting results for only one or two measures of economic inequality is insufficient for policy information affecting economic welfare [5]; however, given these are initial results, the visual impact of the Lorenz affords a useful representation, in the space available, of the inequality under study. To explain our method, we first define the following variables:

$X$ : Utilisation of cancer services
$N$ : Number of LGAs
$i$ : LGA index, $i = 1, 2, \cdots, N$
$I_i$ : Incidence count at $LGA_i$
$T_i$ : Target (normal) separation count at $LGA_i$
$S_i$ : Actual separation count at $LGA_i$
$t_i$ : Actual separation-to-incidence ratio at $LGA_i$

Under normal circumstances, we assume that the number of separations required per new incidence is a constant. Therefore, $T_i = CI_i$ where $C$ is a constant multiplier that can be estimated from hospital records. In reality, the utilisation of cancer services can be affected by many factors, such as cost, travel distance and time, and available modes of transport. To include this effect in the model, the actual separation count, $S_i$, at an LGA is formulated as $S_i = T_i g(D_i) = CI_i g(D_i)$ where $D_i$ is the distance from $LGA_i$ to its nearest hospital; and $g(D_i)$ is a function to model the distance effect. For our simulation, the function $g(D_i)$ chosen is shown in Figure 2. As a measure of utilisation, the actual separation-to-incidence ratio, $t_i$, can be determined as $t_i = \frac{S_i}{I_i} = Cg(D_i); t_i \leq t_{i+1} \forall i$. In reality, chemotherapy is not necessarily prescribed for every cancer. The separation-to-incidence ratio, $t_i$, indicates that for every single cancer incidence, there will be $t_i$ one day chemotherapy separation on average. We do not assume that all cancer incidence is treated with chemotherapy. In practice, if both the $I_i$ and $S_i$ data values are available, they can be used directly in the calculation of $t_i$, without the need of using $g(D_i)$. Once we have the $t_i$ value for each $LGA_i$, the next step is to calculate the coordinates of the Lorenz curve. The calculations are given as follows:

$F_X(t_i)$ : Cumulative proportion of LGAs with utilisation less than $t_i$

$$F_X(t_i) := P(X \leq t_i) = \sum_{j=1}^{i} \frac{1}{N} = \frac{i}{N} \tag{1}$$

$\Phi_X(t_i)$ : Cumulative proportion of utilisation for LGAs with utilisation less than $t_i$

$$\Phi_X(t_i) := \frac{\sum_{j=1}^{i} t_j}{\sum_{j=1}^{N} t_j} \tag{2}$$

$L_X$ : Lorenz curve for $X$

$$L_X := \{(F_X(t_i), \Phi_X(t_i)) : i = 1, 2, \cdots, N\} \tag{3}$$

In the next section, a simulation example will be provided to illustrate the method.

**2. A Simulation Example**

This section presents an example designed based on data that are available in the category of same day chemotherapy. The full set of hypothetical data is provided in Table 1. Thirteen LGAs (A to M) are located linearly with only one hospital in the middle (LGA G) that provides same day chemotherapy service to the people in all LGAs. The population of the LGAs ranges from 10 thousand to over 6 million people, Figure 1. These populations are typical for a regional city and a very large city. Based on the data available from the AIHW [7], the following assumptions are made:

- incidence rate is 529.1 per 100,000 people for all cancers in each LGA.
- the facility located in G has sufficient resources to treat all the patients.
- the estimated ratio of separations to incidence, $C$, is set to a value of 0.6.
- the chemotherapy service is 100% accessible if the patient is located within a distance of 150 km, equivalent to a travel distance of 300 km for each treatment. The accessibility reduces as the patients are farther away from the facility. Distance is used in a loose sense to include all factors that affect accessibility.

The function of the distance factor that impacts the utilisation of chemotherapy is shown in Figure 2. By using Eqs. (1) and (2), the coordinates $F_X(t_i)$ and $\Phi_X(t_i)$ used for plotting the Lorenz curve are given in Table 2. The Lorenz curve showing the distribution of utilisation of chemotherapy among the 13 LGAs is shown in Figure 3. The 45° line represents the ideal perfect equality that all LGAs receive the same level of one day chemotherapy service.

Table 1. Hypothetical simulation data

| LGA | Pop | Location | $D_i$ | $I_i$ | $T_i$ |
|---|---|---|---|---|---|
| A | 10 | (-600, 0) | 1200 | 53 | 32 |
| B | 521 | (-500, 0) | 1000 | 2757 | 1654 |
| C | 5631 | (-400, 0) | 800 | 29794 | 17876 |
| D | 1543 | (-300, 0) | 600 | 8164 | 4898 |
| E | 2054 | (-200, 0) | 400 | 10868 | 6521 |
| F | 5631 | (-100, 0) | 200 | 29794 | 17876 |
| G | 3076 | (0, 0) | 0 | 16275 | 9765 |
| H | 521 | (100, 0) | 200 | 2757 | 1654 |
| I | 4098 | (200, 0) | 400 | 21683 | 13010 |
| J | 4609 | (300, 0) | 600 | 24386 | 14632 |
| K | 521 | (400, 0) | 800 | 2757 | 1654 |
| L | 5631 | (500, 0) | 1000 | 29794 | 17876 |
| M | 6142 | (600, 0) | 1200 | 32497 | 19498 |

Table 2. Data for plotting the Lorenz curve

| $i$ | $LGA_i$ | $F_X(t_i)$ | $t_i$ | Cum $t_i$ | $\Phi_X(t_i)$ |
|---|---|---|---|---|---|
| 1 | M | 0.0769 | 0.0298 | 0.0298 | 0.0093 |
| 2 | A | 0.1538 | 0.0302 | 0.0599 | 0.0188 |
| 3 | B | 0.2308 | 0.0444 | 0.1043 | 0.0328 |
| 4 | L | 0.3077 | 0.0444 | 0.1488 | 0.0467 |
| 5 | C | 0.3846 | 0.0735 | 0.2223 | 0.0698 |
| 6 | K | 0.4615 | 0.0735 | 0.2958 | 0.0929 |
| 7 | D | 0.5385 | 0.1440 | 0.4398 | 0.1381 |
| 8 | J | 0.6154 | 0.1440 | 0.5838 | 0.1834 |
| 9 | E | 0.6923 | 0.4000 | 0.9838 | 0.3090 |
| 10 | I | 0.7692 | 0.4000 | 1.3838 | 0.4346 |
| 11 | F | 0.8462 | 0.6000 | 1.9838 | 0.6231 |
| 12 | G | 0.9231 | 0.6000 | 2.5838 | 0.8115 |
| 13 | H | 1.0000 | 0.6000 | 3.1838 | 1.0000 |

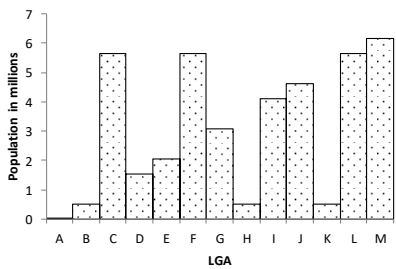

Figure 1. Population of LGAs

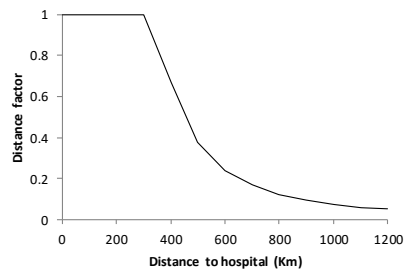

Figure 2. Distance factor

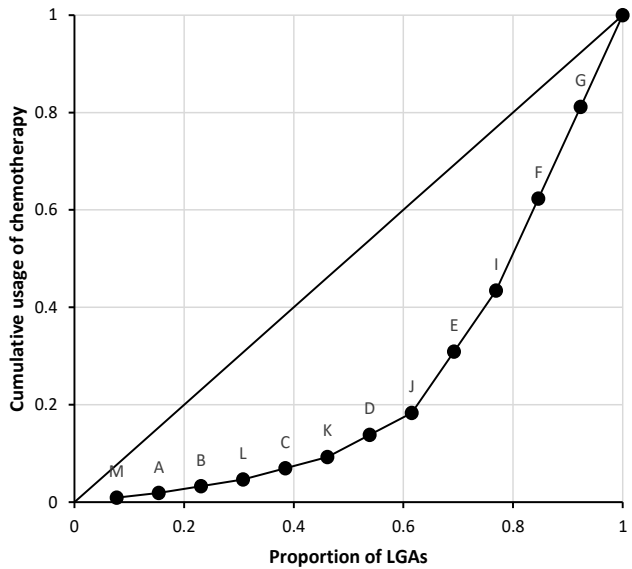

Figure 3. Lorenz curve for the 13 LGAs (A-M)

## 3. Results and Discussion

The example demonstrates how inequality in utilisation of health services is modelled and how the Lorenz curve is constructed form the modelled data. The following observations show that the proposed method accurately reflects service utilisation for the LGAs.

- Three equal sized LGAs with large population {L, C, F} are ranked in ascending order on the Lorenz curve, reflecting travelling distance to the hospital from high to low, and utilisation from low to high.
- Three equal sized LGAs with small populations {B, K, H} shows similar order on the Lorenz curve as above, reflecting similar accessibility and utilisation.
- One large and one small LGAs, equal distanced from the hospital in G, are ranked in the same order, located side by side on the Lorenz curve, reflecting that they both have the same accessibility and utilisation to G. These large and small pairs include {A, M}, {B, L}, {D, J}, and {E, I}.

## 4. Conclusion

Data of separation numbers and incidence rates can be obtained easily in most hospitals. It is practical to use the separation-to-incidence ratio as a basis for measuring inequality in health service accessibility.

The simulation model can be developed further and applied to compare current access of cancer services, or to provide insights for policy makers in future planning by evaluating solutions with alternative or additional facilities.